\UseRawInputEncoding
\documentclass[pra,twocolumn,10pt,superscriptaddress]{revtex4-1}
\usepackage{amssymb,amsmath,amsfonts,xcolor,graphicx}
\usepackage{times}
\usepackage{braket}
\usepackage{float}
\usepackage{lipsum}
\begin{document}

\title{Error-mitigated Geometric Quantum Control over an Oscillator}
\author{Ming-Jie Liang}
\affiliation{Key Laboratory of Atomic and Subatomic Structure and Quantum Control (Ministry of Education),\\ Guangdong Basic Research Center of Excellence for Structure and Fundamental Interactions of Matter,\\ and School of Physics, South China Normal University, Guangzhou 510006, China}
\affiliation{Guangdong Provincial Key Laboratory of Quantum Engineering and Quantum Materials,\\  Guangdong-Hong Kong Joint Laboratory of Quantum Matter,   and Frontier Research Institute for Physics,\\ South China Normal University, Guangzhou 510006, China}

\author{Tao Chen} \email{chentamail@163.com}
\affiliation{Key Laboratory of Atomic and Subatomic Structure and Quantum Control (Ministry of Education),\\ Guangdong Basic Research Center of Excellence for Structure and Fundamental Interactions of Matter,\\ and School of Physics, South China Normal University, Guangzhou 510006, China}

\affiliation{Guangdong Provincial Key Laboratory of Quantum Engineering and Quantum Materials,\\  Guangdong-Hong Kong Joint Laboratory of Quantum Matter,   and Frontier Research Institute for Physics,\\ South China Normal University, Guangzhou 510006, China}

\author{Zheng-Yuan Xue} \email{zyxue83@163.com}
\affiliation{Key Laboratory of Atomic and Subatomic Structure and Quantum Control (Ministry of Education),\\ Guangdong Basic Research Center of Excellence for Structure and Fundamental Interactions of Matter,\\ and School of Physics, South China Normal University, Guangzhou 510006, China}

\affiliation{Guangdong Provincial Key Laboratory of Quantum Engineering and Quantum Materials,\\  Guangdong-Hong Kong Joint Laboratory of Quantum Matter,   and Frontier Research Institute for Physics,\\ South China Normal University, Guangzhou 510006, China}
\affiliation{Hefei National Laboratory,  Hefei 230088, China}

\date{\today}

\begin{abstract}
Quantum information is very fragile to 
environmentally and operationally induced  imperfections. Therefore, the construction of practical quantum computers requires quantum error-correction techniques to protect quantum information. In particular, encoding a logical qubit into the large Hilbert space of an oscillator is a hardware-efficient way of correcting quantum errors. In this strategy, selective number-dependent arbitrary phase (SNAP) gates are vital for  universal quantum control. However, the quality of SNAP gates is considerably limited by the small coupling-induced nonlinearity of the oscillator. 
Here, to resolve this limitation, we propose a robust scheme based on quantum optimal control via  functional theory, by designing an appropriate trajectory for a target operation. Besides, we combine the geometric phase approach with our trajectory design scheme to minimize the decoherence effect, by shortening the gate time. Numerical simulation shows that both errors can be significantly mitigated and that the robustness of the geometric gate against both $X$ and $Z$ errors can be maintained. Therefore, our scheme provides a promising alternative for fault-tolerant quantum computation.

\end{abstract}
\maketitle

\section{Introduction}
Quantum manipulations in quantum systems will inevitably be affected by operational errors and the decoherence effect, which become insurmountable obstacles in practical quantum information processing. To achieve fault-tolerant quantum computation, various quantum error-correction (QEC) protocols have been proposed by encoding quantum information into logical qubits, using the redundancy of a large Hilbert space \cite{Nielsen}. To achieve the goal of quantum information processing with protection, universal control over these logical qubits needs to be realized.  Currently, for a qubit-based quantum computation protocol, logical qubits and QEC are realized by using the surface code \cite{2012-surfacecode,2023-surfacecode-StatePreparation,2023-surfacecode-25}, which is challenging due to the huge physical resource overhead. Compared with the qubit-based strategy, bosonic encoding  provides an alternative route to universal quantum computation, in which quantum information is encoded into an oscillator \cite{GKP, 2021-bosoniccode-review,2021-bosoniccode-review2,2024-GKP} in a hardware-efficient way. Recently, significant progress has been made in this field with a circuit-QED scenario \cite{2016-catcode-lifetime, 2018-bosoniccode-review, 2019-Repetitioncatcode, 2019-binomialcode,2022-catcode, 2023-breakevenpoint}.

In order to realize effective encoding, it is necessary to introduce nonlinearity among bosonic states by dispersively coupling the oscillator to an ancillary qubit. This qubit-oscillator coupled system is an important model  for quantum computation and has been actively explored in various platforms, including trapped ions \cite{ion}, cavity QED \cite{cavity}, circuit QED \cite{circuit}, etc. To process the encoded quantum information, universal control over the coupled system is essential. The selective number-dependent arbitrary phase (SNAP) gate \cite{2015-Uc,2015-SNAP} acts as a powerful way to accumulate the arbitrary phase set on different bosonic states. Together with the displacement gate, arbitrary unitary operations can be constructed, by a sequence of displacement and SNAP gates. Since the displacement gate can be very fast and of high quality, a high-precision SNAP gate plays a vital role for quantum computation with bosonic encoding. In this strategy, a narrow-band pulse at a selected frequency is required to realize a controlled operation. As the interaction-induced nonlinearity of the oscillator is relatively weak, it greatly limits the gate performance. Therefore, various optimized algorithms have been developed to obtained better pulse shapes \cite{2007-QOCFT,2006-QOC1,2013-QOC-QSL,2011-QOC-CRAB,2005-QOC-GRAPE}. Besides, the fidelity of the SNAP gate  \cite{2022-D-SNAP-o} and  the entire unitary operator \cite{2017-Uc-o} are improved by numerical pulse shaping. However, these algorithms often involve solving a complex function to attain high precision, which is hard for large quantum systems. Besides, it is generally difficult to deal further with other intrinsic and/or quantum control imperfections.

Here, to resolve the above difficulties, we propose a scheme using a nonadiabatic geometric phase to implement the SNAP gate with improved quality. First, our scheme is an analytically based one, removing the challenge of a numerical search for large quantum systems. Second, our scheme is fast; the microwave driving field can be strong, comparable to the induced nonlinearity, which removes the narrow-band pulse requirement. Third, our scheme can further be compatible with other optimal-control techniques, as it does not impose any restriction on the pulse shape. By using functional theory, we further propose to optimize the pulse shape to suppress the error from counterrotating terms, which is the main limitation for quantum control in this dispersively coupled system. Meanwhile, the use of geometric phases results in our scheme possessing intrinsic noise-resilience properties. Therefore, our scheme provides a promising strategy for realizing robust SNAP gates, which is essential for universal control over bosonic encoded qubits for quantum computation.

\begin{figure}[tbp]
    \centering
    \includegraphics[width=\columnwidth]{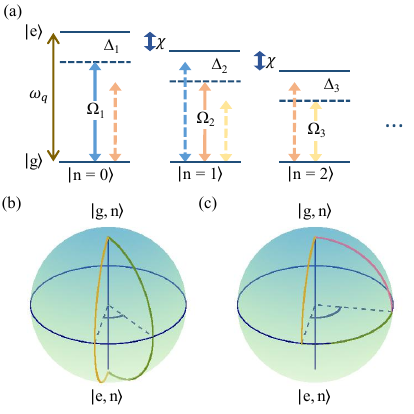}
    \caption{An illustration of our scheme. (a) The energy spectrum of the qubit-cavity dispersively coupled system and the manipulation configuration. The resonant manipulation is obtained  by an external microwave field for specific subspaces, $\{|g,n\rangle, |e,n\rangle\}$ (solid arrows), which also induces off-resonant transitions between neighboring subspaces, $\{|g,n\pm1\rangle, |e,n\pm1\rangle\}$ (dashed arrows). (b),(c) Ideal evolution trajectories for the (b) orange-slice-shaped and (c) path-designed schemes are depicted, with a specific instance in which $\theta_n=\pi/2$. The phase accumulation is proportional to the area enclosed by their corresponding trajectories.  }
    \label{fig}
\end{figure}

\section{Geometric SNAP gate}
Here, we introduce the implementation of the SNAP gate with nonadiabatic geometric phases \cite{AA}, which leads to noise resilience for universal control over the bosonic states in the qubit-oscillator coupled system.

\subsection{System and SNAP gate}

For a qubit-oscillator dispersively coupled system, e.g., a transmon qubit coupled to a three-dimonsional microwave cavity  \cite{2015-SNAP}, the spectrum of the coupled system is shown in Fig. \ref{fig} (a). Assuming $\hbar=1$ hereafter, the system Hamiltonian is
\begin{equation}\label{H0}
    \hat{H}_0=\ \omega_c \hat{a}^\dagger\hat{a}+ \omega_q|e\rangle\langle e|- \chi\hat{a}^\dagger\hat{a}|e\rangle\langle e|,
\end{equation}
where  $\chi$ is the dispersive coupling strength and $\omega_c$,  $\hat{a}$, and $\hat{a}^\dagger$ represent the intrinsic frequency, annihilation, and creation operators of a cavity excitation, respectively. Here, we restrict the qubit to the two-level subspace composed of states $|g\rangle$ and $|e\rangle$, with $\omega_q$ denoting the qubit transition frequency. Due to the dispersive coupling, the qubit transition frequency shifts to $\omega_q-\chi n$ in response to the cavity in an $n$-photon state $|n\rangle$.

We then drive the qubit using a classical microwave field composed of various frequencies. In this case, the interacting Hamiltonian reads \begin{equation}\label{H1}
    \hat{H}_1= \sum_m\frac{\Omega_m(t)}{2} e^{i(\omega_{d,m} t+\phi_m(t))}|g\rangle\langle e|+\mathrm{H.c.},
\end{equation}
where $\Omega_m(t)$, $\omega_{d,m}$, and $\phi_m(t)$ are the amplitude, frequency, and phase of the microwave field, respectively.
The driving frequency $\omega_{d,m}$ is targeted to coupled the transition of the subspace with the Fock state $|m\rangle$, i.e., $\omega_{d,m}=\omega_q-\Delta_m-\chi m$. Then, after transforming to the interaction picture, under the condition of $|\Omega_m|\ll\chi$, the interaction Hamiltonian in the $n$th subspace reads
\begin{eqnarray}\label{HR}
\hat{H}_{n} = \frac{\Delta_n }{2}\hat{\sigma}^z_n+\frac{\Omega_n(t)}{2}\left[\cos\phi_n(t)\hat{\sigma}^x_n+\sin\phi_n(t)\hat{\sigma}^y_n\right],
\end{eqnarray}
where $\hat{\sigma}^j_n$, with $j\in \{x,y,z\}$, denotes the Pauli operators acting on the subspace spanned by $\{|g,n\rangle,|e,n\rangle\}$, as shown in Fig. \ref{fig} (a). Thus, this interaction allows for independent control of each subspace of the qubit-oscillator coupled system.

With this selective addressing,
the SNAP gate,
\begin{eqnarray} \label{snap}
\hat{S} (\vec{\theta})=\prod^\infty_{n=0}\hat{S}_n\left(\theta_n\right) =\sum^\infty_{n=0}e^{i\theta_n}|n\rangle\langle n|,
\end{eqnarray}
with $\vec \theta=\{\theta_n\}^{\infty}_{n=0}$, can be implemented, using the two-segment scheme \cite{ngqc}, as shown in Fig. \ref{fig} (b). The scheme is based on the Hamiltonian in Eq. (\ref{HR}), where a geometric phase is induced for each subspace following a cyclic evolution with a period of $T_O$. When $\Delta_n=0$, the pulse areas and phases should meet the following conditions:
\begin{eqnarray}\label{Two-segments}
\int_{0}^{T_O/2}\Omega_n\left(t\right)dt &=& \pi,\quad \phi_n(t)=-\theta_n-\frac{\pi}{2}, \quad t\in\left[0,T/2\right),\nonumber\\
\int_{T_O/2}^{T_O}\Omega_n\left(t\right)dt &=& \pi,\quad \phi_n(t)=\frac{\pi}{2},\quad t\in\left[T/2,T\right].
\end{eqnarray}

However, the operation time for systems with smaller amplitudes will be disproportionately prolonged and thus the decoherence effect will become the dominant error source. To achieve a faster gate implementation, the condition $|\Omega_m|\ll\chi$ needs to be removed or mitigated. In this case, it is crucial to  take the effect of the counterrotating terms \cite{2023-Fast} into consideration, i.e., Eq. Eq. (\ref{HR}) becomes
\begin{eqnarray}\label{Hnm}
    \hat{H}_{n,m}&=&\frac{\Delta_n }{2}\hat{\sigma}^z_n+\frac{\Omega_n(t)}{2} e^{i\phi_n(t)}|g,n\rangle\langle e,n|\\     && +
    \sum_{m\neq n}\frac{\Omega_m(t)}{2}e^{i((n-m) \chi t+\phi_m(t))}|g,n\rangle\langle e,n| +\mathrm{H.c.}\nonumber
\end{eqnarray}
The second line contains the counterrotating terms, which introduce error that affects the accuracy of the operation. We then need to strike a balance between decoherence and operational errors to maintain the desired gate performance.

\subsection{Geometric SNAP gate with pulse shaping}

Since there is no restriction on the pulse shapes in the above scheme, here we propose to incorporate the pulse-shaping technique into the implementation. We first focus on the single-frequency case in Eq. (\ref{HR}) and then consider many frequencies. To begin with, we introduce a set of auxiliary bases $|Q_{n,u}(t)\rangle=e^{if_{n,u}(t)}|q_{n,u}(t)\rangle$, and
\begin{eqnarray}
|q_{n,+}(t)\rangle &=& \cos\frac{\zeta_n(t)}{2}|g,n\rangle+\sin\frac{\zeta_n(t)}{2}e^{i\xi_n(t)}|e,n\rangle,\nonumber\\
|q_{n,-}(t)\rangle &=& \sin\frac{\zeta_n(t)}{2}e^{-i\xi_n(t)}|g,n\rangle-\cos\frac{\zeta_n(t)}{2}|e,n\rangle,
\end{eqnarray}
with $u\in\{+,-\}$, $f_{n,u}(0)=0$, the variables $\zeta_n(t)$ and $\xi_n(t)$ representing the polar and azimuth angles, respectively, that determine the  evolution trajectories on the Bloch sphere. The correspondence between the evolution trajectories and the relevant parameters of $\hat{H}_{n}$ can be derived by solving the Schr\"{o}dinger equation, $i\partial|Q_{n,\pm}(t)\rangle/\partial t=\hat{H}_{n}|Q_{n,\pm}(t)\rangle$ and are
\begin{eqnarray}    \dot{\zeta}_n(t)&=&\Omega_n(t)\sin(\phi_n(t)-\xi_n(t)),\nonumber\\
    \dot{\xi}_n(t)&=&-\Delta_n(t)-\Omega_n(t)\cot\zeta(t)\cos(\phi_n(t)-\xi_n(t)),\nonumber\\
    f_{n,+}(t)&=&-f_{n,-}(t)\nonumber\\
    &=&-\frac{1}{2}\int^t_0\frac{\dot{\xi}_n(t')[\cos\zeta_n(t')-1]-\Delta_n(t')}{\cos\zeta_n(t')}dt'.
\end{eqnarray}
In essence, distinct trajectories on the Bloch sphere are associated with varying the shape of these time-dependent parameters, which, in turn, determine the gate performance.

When the evolution states are set to satisfy the cyclic condition of $U(T)\prod^\infty_{n=0}|Q_{n,\pm}(0)\rangle=\prod^\infty_{n=0}e^{if_{\pm}(T)}|Q_{n,\pm}(0)\rangle$ with $|Q_{n,\pm}(T)\rangle=|Q_{n,\pm}(0)\rangle$, after a cyclic evolution with the time $T$, the corresponding evolution operator $U(T)$ is
\begin{eqnarray}
    U(T)&=&\prod^\infty_{n=0}\sum_{u\in\{+,-\}}e^{if_{n,u}(T)}|Q_{n,u}(0)\rangle\langle Q_{n,u}(0)|\nonumber\\
    &=&\prod^\infty_{n=0}[\cos\gamma_n+i\sin\gamma_n\cos\zeta_n(0)\hat{\sigma}^z_n\nonumber\\
    &&+i\sin\gamma_n\sin\zeta_n(0)(\cos\xi(0)\hat{\sigma}^x_n+\sin\xi(0)\hat{\sigma}^y_n)]\nonumber\\    &=&\prod^\infty_{n=0}e^{i\gamma_n\vec{\upsilon}_n\cdot\vec{\sigma}_n},
\end{eqnarray}
where $\gamma_n=f_{n,+}=-f_{n,-}$ is the total accumulated phase, $\vec{\upsilon}_n=(\sin\zeta_n(0)\cos\xi_n(0),\sin\zeta_n(0)\sin\xi_n(0),\cos\zeta_n(0))$, and $\vec{\sigma}_n=(\hat{\sigma}^x_n,\hat{\sigma}^y_n,\hat{\sigma}^z_n)$. This allows for independent control of each subspace $\{|g,n\rangle,|e,n\rangle\}$ in the cavity.

The total phase accumulation $\gamma_n$ can be decomposed into dynamic $\gamma^d_n$ and geometric $\gamma^g_n$ components, which are
\begin{eqnarray}
    &&\gamma^d_n=\frac{1}{2}\int^T_0\frac{\dot{\xi}_n(t)\sin^2\zeta_n(t)+\Delta_n(t)}{\cos\zeta_n(t)}dt,\nonumber\\
    &&\gamma^g_n=-\frac{1}{2}\int^T_0\dot{\xi}_n(t)[1-\cos\zeta_n(t)]dt.
\end{eqnarray}
To ensure that a geometric SNAP gate is implemented, it is essential to set the accumulated dynamical phase to be zero, i.e., $\gamma^d_n = 0$. This condition can be satisfied by appropriate design of the evolution path.
The standard implementation employs a two-segment scheme \cite{ngqc}, where $\gamma^g_n=\theta_n$ and $\gamma^d_n=0$. To minimize the decoherence effect, the qubits are initialized in the ground state $|g\rangle$, corresponding to the initial condition of $(\zeta_n(0),\xi_n(0))=(0,0)$, while allowing the cavity to be prepared in any arbitrary state.  Following this cyclic evolution, obtained through the application of multiple frequency driving,  the SNAP gate in Eq. (\ref{snap}) can be realized.

\section{QOC via Functional Theory}\label{qoc}

Note that, in the above construction, when
$|\Omega_m| \ll \chi$ is not met, the counterrotating terms in Eq. (\ref{Hnm}) introduce non-negligible gate errors. A pulse-optimization scheme to suppress this error has recently been proposed in Ref. \cite{2023-Fast}. However, this approach is limited to the standard two-segment scheme. Here, we aim to expand the framework to unset trajectories to further enhance the SNAP-gate performance. Therefore, we mitigate errors through pulses shaping based on the quantum optimal control via functional theory (QOCF) approach. This method is designed to resist errors throughout the entire evolution of the SNAP gate.

QOCF provides an efficient means to derive controllable equations of an optimal pulse that conform to the variation of a suitably defined function.
The controllable equations that we consider here have two parts and our goal is to find a suitable pulse shape to minimize the sum of the two. The first part is the distance between the target state $|\psi_{tar}\rangle$ and the final state $|\psi(T)\rangle$ \cite{2001-QOCFT-J1}, i.e.,
\begin{eqnarray}\label{J1}
J_1=  \||\psi(T)\rangle-|\psi_{tar}\rangle\|^2,
\end{eqnarray}
which determines the process gate fidelity. As the evolution state $|\psi(t)\rangle$ has to satisfy the time-dependent Schr\"{o}dinger equation \cite{1988-QOCFT-J2,1998-QOCFT-J2}, the second part is the derivation from the target trajectory set by the Schr\"{o}dinger equation, i.e.,
\begin{eqnarray}
J_2&=&2\text{Im}\int^T_0dt\langle\lambda(t)|(i\frac{\partial}{\partial t}-\hat{H}_{n,m}(t))|\psi(t)\rangle,
\end{eqnarray}
where $|\lambda(t)\rangle$ are the Lagrange multipliers and
\text{Im}(...) donates the imaginary part of the function. Finally, we obtain the Lagrange function as
\begin{eqnarray} \label{Lagrange} J(|\lambda(t)\rangle,|\psi(t)\rangle,\Omega_m(t),\Delta_n(t))=J_1+J_2
\end{eqnarray}
and we can formulate this function as a standard optimal-control problem to address the obstacles encountered in implementing the SNAP gate. In this case, our goal is to find a pulse that meets the following conditions:
\begin{eqnarray}\label{deltaJ}
    \delta J=0,\quad\delta_\psi J =\delta_\lambda J=\delta_{\Omega_m} J=\delta_{\Delta_n} J=0.
\end{eqnarray}
Here we have omitted the derivatives with respect to the  conjugate variables $\langle\lambda(t)|$ and $\langle\psi(t)|$, as the necessary conditions derived from them are the same as Eq. (\ref{deltaJ}).

\begin{figure}[tbp]
    \centering
\includegraphics[width=\columnwidth]{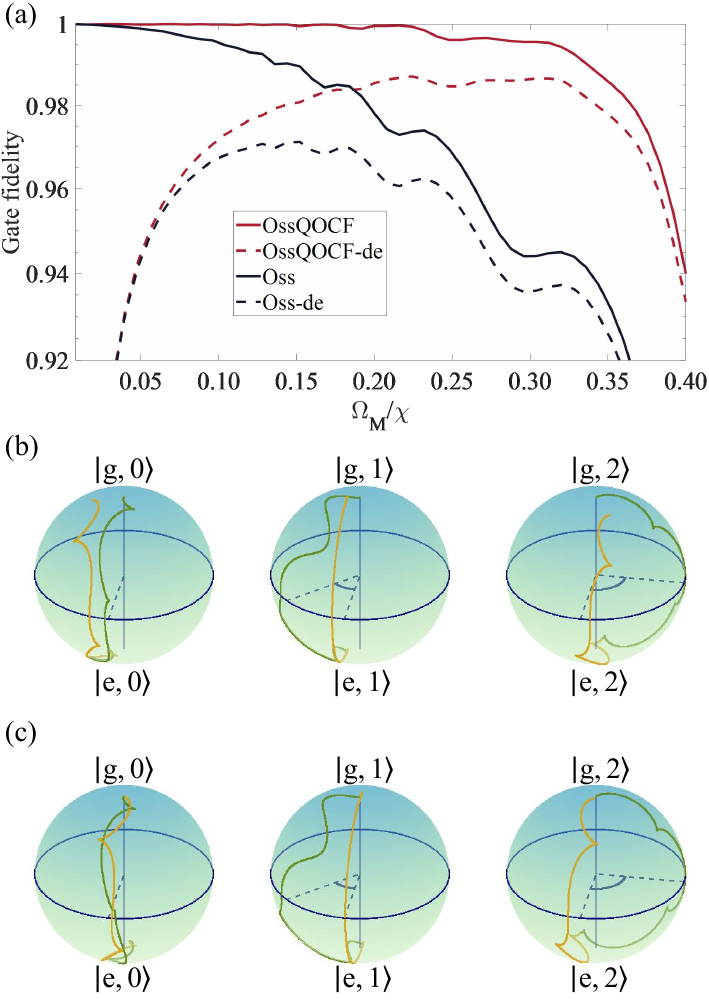}
\caption{The performance of the orange-slice-shaped (Oss) implementation of the SNAP gate of $\vec\theta=(0,-\pi/4,\pi/2)$ with the counterrotating terms. (a) The SNAP-gate fidelity as a function of $\Omega_M/\chi$ (solid lines) without and (dashed lines) with decoherence. Compared with the unoptimized scheme (blue lines), our optimized scheme (red lines) can maintain higher gate fidelity for larger $|\Omega_m|$, by mitigating the error due to the counterrotating terms.
A schematic illustration of the evolution paths  in the subspace $\{|g,n\rangle,|e,n\rangle\}$  (b) without and (c) with the QOCF techniques.}
    \label{fig1}
\end{figure}

Next, we focus on the full interaction Hamiltonian  in  Eq. (\ref{Hnm}): the evolution state $|\psi_n(t)\rangle$  can then be expressed as $|\psi_n(t)\rangle=c_{g,n}|g,n\rangle+c_{e,n}|e,n\rangle$, where $c_{g,n}$ and $c_{e,n}$ represent the complex amplitude of states $|g,n\rangle$ and $|e,n\rangle$, respectively. With the initial condition of $|\psi_n(0)\rangle=|g,n\rangle$, our target state will be $|\psi_{n,tag}\rangle=e^{i\theta_n}|g,n\rangle$.Under these settings, we derive the restriction as
\begin{eqnarray}\label{restrictions}
&&\langle(i\frac{\partial}{\partial t}-\hat{H}_{n,m}(t))\psi_n(t)|\delta\lambda_n(t)\rangle=0,|\psi_n(0)\rangle,\nonumber\\
&&\langle(i\frac{\partial}{\partial t}-\hat{H}_{n,m}(t))\lambda_n(t)|\delta\psi_n(t)\rangle=0, |\lambda_n(T)\rangle=|\psi_{n,tag}\rangle,\nonumber\\
&&\text{Im}\langle\lambda_n(t)|\frac{\delta\hat{H}_{n,m}}{\delta\Omega_m}\delta\Omega_m|\psi_n(t)\rangle=0,\nonumber\\    &&\text{Im}\langle\lambda_n(t)|\frac{\delta\hat{H}_{n,m}}{\delta\Delta_n}\delta\Delta_n|\psi_n(t)\rangle=0.
\end{eqnarray}
The choice of $\Omega_m$ and $\Delta_n$ that strictly meet the above equations will ensure that Eq. (\ref{deltaJ}) is satisfied. Specifically, in the following, a tailored pulse shape is identified that mitigates the impact of the counterrotating terms. For simplicity,  the primary consideration is the lower-frequency terms with $|m-n|\leq 1$, while the higher-order counterrotating terms ($|m-n|>1$)  are neglected. As detailed in Appendix \ref{constrain}, the relationship between $\Delta_n$ and $\Omega_m$ is found to be
\begin{eqnarray}\label{relationship}
    \Delta_n=-i\Xi_n\cot(\Xi_n(t-T)),
\end{eqnarray}
where $\Xi_n=\sqrt{\Delta^2_n+\Pi_n\Pi^*_n}$, $\Pi_0=\Omega_{0}e^{i\phi_{0}}+\Omega_{1}e^{i(\phi_{1}-\chi t)}$, and $\Pi_n= \Omega_{n-1}e^{i(\phi_{n-1}-\chi t)}+\Omega_{n}e^{i\phi_{n}}+\Omega_{n+1}e^{i(\phi_{n+1}+\chi t)} $ for $n\geq 1$.

\begin{figure}[tbp]
    \centering
    \includegraphics[width=\columnwidth]{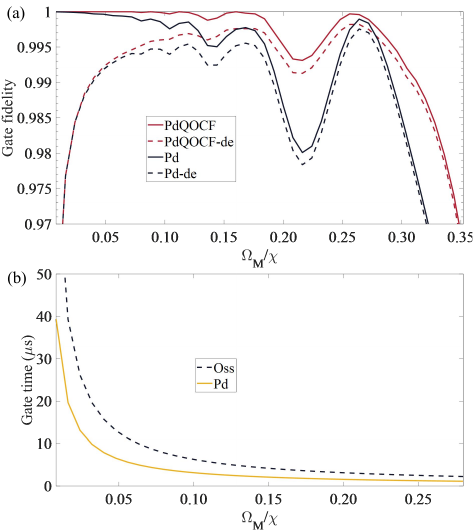}
\caption{The performance for the $\vec\theta=(0,-\pi/4,\pi/2)$ gate as a function of $\Omega_M/\chi$, with counterrotating terms. (a) The SNAP-gate fidelity for the path-designed (Pd) scheme without and with the decoherence are plotted as solid and dashed lines, respectively.
(b) The gate times for the orange-slice-shaped scheme and our path-designed scheme.}
    \label{fig3}
\end{figure}

We then take the $\vec{\theta}=(0,-\pi/4,\pi/2)$ gate as an example: the same pulse shape is chosen as $\Omega_m=\Omega_M\sin(\pi t/\tau)$, with $\tau$ being the evolution time for each segment. We use a slight adjustment in the pulse shape to ensure that Eq. (\ref{relationship}) approximately satisfied (for details, see Appendix \ref{adjust}). Besides, this gate is only a linear correlation one, not a special case (for details, see Appendix \ref{correlation}).
The fidelity of the SNAP gate as a function of  $\Omega_M/\chi$ is shown in Fig. \ref{fig1}(a), with the dispersive coupling strength $\chi=2\pi\times2.5$ MHz and the qubit decay and dephasing rates being $2\pi\times1.45$ kHz. As the decoherence for  a cavity is much smaller  \cite{2013-C-ka}, it is neglected.  In the limiting case of $|\Omega_m|\ll\chi$, the error induced by the counterrotating terms is negligibly small. However, decoherence will seriously impair the gate fidelity, as the gate time is longer for smaller $\Omega_M/\chi$. On the other hand, while $\Omega_M/\chi$ is becoming larger, $|\Omega_m|\ll\chi$ will finally be invalid, and error induced by the counterrotating terms will gradually arise and result in larger distortion of the gate. However, for the QOCF approach, the amplitude of the pulse in the intermediate and trailing portions can still maintain a high gate fidelity.

We also illustrate the trajectories of the SNAP gate. Compared to the ideal path depicted in Fig. \ref{fig}(b),  under the influence of the counterrotating terms, the evolution state is more tortuous and fails to return to the initial starting point under the influence of the counterrotating terms, as shown in Fig. \ref{fig1}(b). However, after applying the proposed QOCF techniques, the evolution state can still approximately go back to the original starting point, as shown in Fig. \ref{fig1}(c), and thus it can lead to higher gate fidelity.

\section{Path-designed control}

In addition to designing the pulse shape, the evolution loop on the Bloch sphere can also be reconstructed, to further shorten the gate time. This can be applied to the SNAP gate and can reduce the decoherence error.
As shown in Fig. \ref{fig}(c), we adopt a three-segment scheme \cite{NGQC-SP2-O*}, where the cyclic evolution trajectories are formed by two longitudinal lines and one latitudinal line. Similarly, for each driving frequency, we consider the Hamiltonian in Eq. (\ref{HR}). The pulse areas and phases of the three segments should meet the following conditions in each subspace:
\begin{subequations}\label{po-condition}
\begin{eqnarray}
\int^{T_{1,n}}_0\Omega_n(t)dt&=&\Lambda_n,\quad \Delta_n=0,\nonumber\\
    \phi_n(t)&=&\xi^b_n+\frac{\pi}{2},\qquad t\in[0,T_{1,n});
\end{eqnarray}
\begin{eqnarray}
\int^{T_{2,n}}_{T_{1,n}}\Omega_n(t)dt&=&|\kappa_n\tan\Lambda_n\cos^2\Lambda_n| \nonumber\\
\Delta_n&=&\frac{\kappa_n\sin^2\Lambda_n}{T_{2,n}-T_{1,n}},\nonumber\\
    \phi_n(t)&=&\left\{\begin{array}{c}
         \xi_n(t)+\pi, \quad 0<\Lambda_n<\pi/2 \\
         \xi_n(t), \quad \pi/2<\Lambda_n\leq\pi
    \end{array}\right. \nonumber\\
   \int^{T_{2,n}}_{T_{1,n}}\dot{\xi}(t)&=&-\kappa_n,\quad t\in[T_{1,n},T_{2,n});
\end{eqnarray}
    \begin{eqnarray}
\int^{T_{P}}_{T_{2,n}}\Omega_n(t)dt&=&\Lambda_n,\quad \Delta_n=0,\nonumber\\
    \phi_n(t)&=& \xi^a_n-\pi/2, \quad t\in[T_{2,n},T_{P}],
\end{eqnarray}
\end{subequations}
where $\Lambda_n$ is a constant of the accumulated polar angle in the first and third segments and $\Lambda_n\in(0,\pi/2)\cup(\pi/2,\pi]$, $\kappa_n=\xi^b_n-\xi^a_n$, $\xi^a_n=\xi_n(0)$, $\gamma^g_n=\kappa_n(1-\cos\Lambda_n)/2=\theta_n$, and $\gamma^d_n=0$.
Similarly, the proposed quantum optimal-control method can still be combined, by setting the Hamiltonian for each segment to fulfill the requirement in Eq. (\ref{relationship}), to mitigate the effect due to counterrotating terms.

Next, we also take the $\vec\theta=(0,-\pi/4,\pi/2)$ gate as an example, to compare the performance of the optimized path scheme with the orange-slice-shaped scheme. The system parameters are the same as those in Fig. \ref{fig1}, the gate time $T_P$ is defined as a function of $\Lambda_0$ and $\gamma^g_0$, and we take $\Lambda_0=0.501\times\pi$ for a shorter gate time $T_P$. The gate performance is shown in Fig. \ref{fig3}(a). For the same pulse amplitude, the path-designed scheme exhibits superior performance compared to the orange-slice-shaped scheme. Moreover, it mitigates the significant detrimental effect of decoherence due to the excessively long duration of the pulse at the front part. We also present the result of the path-designed scheme with the QOCF approach in Fig. \ref{fig3}(a), demonstrating notable resistance to decoherence and counterrotating effects over a wide range of $\Omega_M$ selections.
In addition, the comparison of the gate times is shown in Fig. \ref{fig3}(b), the time for the path-designed scheme being approximately half of that for the orange-slice-shaped scheme. Thus, our scheme speeds up the implementation of the SNAP gate. Here, we choose the three-segment trajectory scheme to demonstrate its effectiveness in reducing the gate time, because increasing the number of segments beyond three will not lead to significant further acceleration of the gate time and only offers a slight improvement in robustness. 

\begin{figure}[tbp]
    \centering
    \includegraphics[width=\columnwidth]{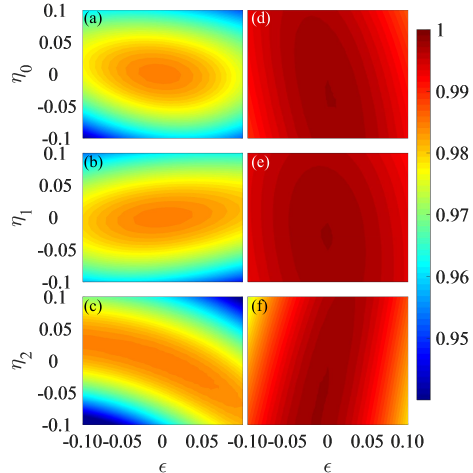}
\caption{The gate robustness with respect to the deviation fraction $\epsilon$ of the driving amplitude and the qubit-frequency drift fraction $\eta_n$ in subspace $\{|g,n\rangle,|e,n\rangle\}$. The gate robustness of the SNAP gate for (a)-(c) the orange-slice-shaped scheme and (d)-(f) for the path-designed scheme with QOCF techniques.}
    \label{fig4}
\end{figure}

Due to the constraints of the implementation, the operational control error and the dephasing noise resulting from the qubit-frequency drift also need to be taken into account, i.e. the $X$ and $Z$ errors. As the path-designed scheme inherits the intrinsic merit of geometric phases from the orange-slice-shaped scheme, its exhibits robustness against the implementation of the SNAP gate within a certain error range. Under the influence of both $X$ and $Z$ errors, the optimal pulse is changed to $\Omega_E=(1+\epsilon)\Omega'$ and $\Delta_{n,E}=\Delta'_n-\eta_n\Omega_M$, respectively. We numerically show the effectiveness of the path-designed scheme with QOCF and the orange-slice-shaped scheme in Fig. \ref{fig4}, with $\Omega_M=2\pi\times0.66$ MHz, $\Omega_{m, \varepsilon}/\Omega_M=-0.1$ and $\Delta_{n, \varepsilon}/\Omega_M=\{-0.01,0.00,0.01\}$. It is shown that the path-designed scheme will have a better  performance.

\section{Discussion and conclusions}\label{Higher order}
Besides the considered main error source of the counterrotating terms, our optimized scheme can further be incorporated to compensate for higher-order effects (for details, see Appendix \ref{highorder}).
Further refinements of our method can be considered as follows. First, in the QOCF approach, ensuring the function in Eq. (\ref{relationship}) rigorously will yield higher fidelity. For our path-designed scheme, the trajectory can be designed as a circle \cite{2024-Circle}, potentially further reducing the gate time. However, this comes with more stringent requirements in terms of physical implementation.
Second, in combination with the displacement operator, the gate set $\{S(\vec\theta),D(\alpha)\}$ provides universal control over the cavity \cite{2015-Uc,2022-D-SNAP-o}. Different values of $\vec\theta$ correspond to varying degrees of optimization. For example, choosing $\vec\theta=(\pi,0,0,0,0,0)$ may mitigate counterrotating terms due to its inherent symmetry. Finally, crosstalk remains a concern in other studies,  such as in transmon-qubit systems with parametrically tunable coupling \cite{2018-PTC-1,2018-PTC-2}: the QOCF approach can also be applied to overcome crosstalk.

In conclusion, we propose a scheme with the QOCF approach and a path-designed scheme to suppress the effect of counterrotating terms and decoherence that existed in the SNAP gate.  Numerical simulations demonstrate that in a wide range of pulse amplitude selections, both decoherence and counterrotating terms can be significantly mitigated. Besides, the intrinsic merit of geometric phases enhances the robustness of the SNAP gate against the $X$ and $Z$ errors.

\acknowledgements

This work was supported by the National Natural Science Foundation of China (Grants No. 12275090 and No. 12305019), the Guangdong Provincial Quantum Science Strategic Initiative (Grant No. GDZX2203001), and the Innovation Program for Quantum Science and Technology (Grant No. 2021ZD0302303).

\section*{Date availablity}
The data that support the findings of this paper are
not publicly available because they contain commercially sensitive information. The data are available from the authors upon reasonable request.


\appendix

\section{Derivation of the QOCF constraint}\label{constrain}
Here, we present on the derivation of the QOCF constraints from the Hamiltonian presented in Eq. (\ref{Hnm}). For simplicity, the objective of determining the minimum value of $J_1$ can be transformed as
\begin{widetext}
\begin{eqnarray}
    \min _{\Omega_m(t), \Delta_n(t)} J_1&=& \min_{\Omega_m(t), \Delta_n(t)} \||\psi_n(T)\rangle-|\psi_{n,tar}\rangle\|^2 
    =\max_{\Omega_m(t), \Delta_n(t)}2\text{Re}\langle\psi_n(T)|\psi_{n,tar}\rangle.
\end{eqnarray}
The Lagrange function in Eq. (\ref{Lagrange}) can be expressed in the form of
\begin{eqnarray}
J &=& 2\text{Re}\langle\psi_n(T)|\psi_{n,tar}\rangle
+2\text{Im}\int^T_0dt\langle\lambda_n(t)|(i\frac{\partial}{\partial t}-\hat{H}_{n,m}(t))|\psi_n(t)\rangle,\\
&=&\langle\psi_n(T)|\psi_{n,tar}\rangle+\langle\psi_{n,tar}|\psi_n(T)\rangle 
-i\int^T_0dt\left[\langle\lambda_n(t)|(i\frac{\partial}{\partial t}-\hat{H}_{n,m}(t))\psi_n(t)\rangle
-\langle(i\frac{\partial}{\partial t}-\hat{H}_{n,m}(t))\psi_n(t)|\lambda_n(t)\rangle\right].\nonumber
\end{eqnarray}
Next, we take the Fr$\mathrm{\acute{e}}$chet derivatives of $J$ with respect to independent variables of $|\lambda_n(t)\rangle,|\psi_n(t)\rangle$, $\Omega_m(t)$, and $\Delta_n(t)$ and the result reads
\begin{eqnarray}
\delta_{|\lambda_n\rangle} J&=& i\int^T_0dt\langle(i\frac{\partial}{\partial t}-\hat{H}_{n,m}(t))\psi_n(t)|\delta\lambda_n(t)\rangle=0,\nonumber\\
 \delta_{|\psi_n\rangle} J&=&\langle\psi_{n,tar}|\delta\psi_n(t)\rangle\bigg|^T
 -i\int^T_0dt\langle(i\frac{\partial}{\partial t}-\hat{H}_{n,m}(t))\lambda_n(t)|\delta\psi_n(t)\rangle
+\langle\lambda_n(t)|\delta\psi_n(t)\rangle\bigg|^T_0=0,\nonumber   \\
    \delta_{\Omega_m}J &=& -2\text{Im}\int^T_0dt\langle\lambda_n(t)|\frac{\delta\hat{H}_{n,m}}{\delta\Omega_m}\delta\Omega_m|\psi_n(t)\rangle=0,\nonumber\\
\delta_{\Delta_n}J &=& -2\text{Im}\int^T_0dt\langle\lambda_n(t)|\frac{\delta\hat{H}_{n,m}}{\delta\Delta_n}\delta\Delta_n|\psi_n(t)\rangle=0,
\end{eqnarray}
and then the restriction in Eq. (\ref{restrictions}) can be obtained.
\end{widetext}

After that, we substitute the Hamiltonian in Eq. (\ref{Hnm}) for further calculations. Here, the Lagrange multipliers are defined as $|\lambda_n(t)\rangle=\lambda_{1,n}|g,n\rangle+\lambda_{2,n}|e,n\rangle$, which leads to
\begin{eqnarray}\label{A5a}
i\frac{\partial\lambda_{1,n}}{\partial t}-\Pi^*_n\lambda_{2,n}-\Delta_n\lambda_{1,n}=0,\nonumber\\
i\frac{\partial\lambda_{2,n}}{\partial t}-\Pi_n\lambda_{1,n}+\Delta_n\lambda_{2,n}=0,
\end{eqnarray}
and
\begin{eqnarray}\label{A5b}     &&\text{Im}\int^T_0dt\left[\lambda^*_{1,n}\Pi^*_nc_{e,n}+\lambda^*_{2,n}\Pi_nc_{g,n}\right]=0 (\Omega_m\neq0),\nonumber\\
&&\text{Im}\int^T_0dt\left[\lambda^*_{1,n}c_{g,n}-\lambda^*_{2,n}c_{e,n}\right]=0, (\Delta_n\neq0).
\end{eqnarray}
Thus, it can be demonstrated that 
\begin{eqnarray}
    \Pi_0&=& \Omega_{0}e^{i\phi_{0}}+\Omega_{1}e^{i(\phi_{1}-\chi t)},\nonumber\\
    \Pi_n&=& \Omega_{n-1}e^{i(\phi_{n-1}-\chi t)}+\Omega_{n}e^{i\phi_{n}} \nonumber\\
    &&+\Omega_{n+1}e^{i(\phi_{n+1}+\chi t)} \quad(n\geq 1).
\end{eqnarray}
Since the Lagrange multipliers are required to satisfy the condition $|\lambda_n(T)\rangle=|\psi_{n,tag}\rangle=e^{i\theta_n}|g,n\rangle$, we assume that the solutions can be expressed as a power series with respect to the variable time $t'$. At $t'=T$, they can be expanded as
\begin{eqnarray}
    \lambda_{1,n}(t)=\sum^\infty_{k=0}a_{k,n}(t'-T)^k,\nonumber\\
    \lambda_{2,n}(t)=\sum^\infty_{k=0}b_{k,n}(t'-T)^k,
\end{eqnarray}
where $a_{k,n}$ and $b_{k,n}$ are complex parameters. It is obvious that $a_0=e^{i\theta_n}$, $b_0=0$. By substituting the power-series solutions into Eq. (\ref{A5a}), the coefficients for the remaining terms can be determined. The recursion relationship is
\begin{eqnarray}
    a_{k+1,n}&=&\frac{-i}{2(k+1)}(\Pi^*_kb_k+\Delta_ka_k),\nonumber\\
    b_{k+1,n}&=&\frac{-i}{2(k+1)}(\Pi_ka_k-\Delta_kb_k),
\end{eqnarray}
the coefficients for the remaining terms in the power-series being
\begin{eqnarray}
    a_n=\left\{\begin{array}{c}
         \frac{(-i)^ne^{i\theta_n}}{2^nn!} \Delta_n(\Delta^2_n+\Pi_n\Pi^*_n)^{\frac{n-1}{2}} \quad \text{n odd} \\
         \\
         \frac{(-i)^ne^{i\theta_n}}{2^nn!} (\Delta^2_n+\Pi_n\Pi^*_n)^{\frac{n}{2}} \quad \text{n even}
    \end{array}\right.,\nonumber\\
    b_n=\left\{\begin{array}{c}
         \frac{(-i)^ne^{i\theta_n}}{2^nn!} \Pi_n(\Delta^2_n+\Pi_n\Pi^*_n)^{\frac{n-1}{2}} \quad \text{n odd} \\
         \\
         0 \quad \text{n even}
    \end{array}\right..
\end{eqnarray}
The power-series solutions for the Lagrange multipliers can be expressed as
\begin{eqnarray}\label{A10}
    \lambda_{1,n} &=& e^{i\theta_n}\left[\cos(\frac{\Xi_n}{2}(t'-T))-i\frac{\Delta_n}{\Xi}\sin(\frac{\Xi_n}{2}(t'-T))\right],\nonumber\\
    \lambda_{2,n} &=& -ie^{i\theta_n}\frac{\Pi_n}{\Xi_n}\sin(\frac{\Xi_n}{2}(t'-T)),
\end{eqnarray}
where $\Xi_n=\sqrt{\Delta^2_n+\Pi_n\Pi^*_n}$.  Finally, the constraint relationship can be obtained by combining Eq. (\ref{A10}) and Eq. (\ref{A5b}),  as Eq. (\ref{relationship}) in the main text. Thus, when $\Omega_m$ and $\Delta_n$ strictly satisfy the above equations, an optimized gate fidelity can be achieved.

\section{Linear adjustment of the parameters} \label{adjust}
The constraint in Eq. (\ref{relationship}) requires complex pulse-shape modulation. To simplify the complexity, we make an approximation that aims to satisfy the constraint relationships as closely as possible. At $t'=T$, we perform a power-series expansion of the constraint relationships as follows:
\begin{eqnarray}\label{appoxrelationship}
    \Delta_n&=&-i\Xi\cot(2\Xi(t'-T))\nonumber\\
    &=&-i\Xi\left[\frac{1}{2\Xi(t'-T)}-\frac{2\Xi(t'-T)}{3}\cdots\right].
\end{eqnarray}
Here, we introduce a slight linear adjustment to facilitate the integration with the path-designed control scheme.  That is, we set $\Omega_m'=(\Omega_M+\Omega_{m, \varepsilon})\sin(\pi t/\tau)$ and $\Delta_n'=\Delta_n+ \Delta_{n, \varepsilon}$ to approximately satisfy Eq. (\ref{appoxrelationship}), where $\Omega_{m, \varepsilon}$ and $\Delta_{n, \varepsilon}$ are real parameters and they are within the $\pm 10\%$ range of   $\Omega_{M}$ and $\Delta_{n}$, respectively.  Finally, the optimal parameters are determined through numerical optimization. In contrast to global optimization, the optimization parameters for each subspace can be determined through the  constraint relationships in Eq. (\ref{relationship}). This eliminates the need for a global search, thereby significantly reducing the required computational resources.

\section{Correlation analysis of different SNAP gates}\label{correlation}
Different values of $\vec\theta$ correspond to varying degrees of optimization. This can be explained through the correlation between the constraint relationships of different subspaces. Here, we take $\Xi^2$ within the constraint relationships as a  measure of the correlation between the different subspaces. It reads
\begin{eqnarray}
\Xi_n^2&=&\Delta^2_n+\Pi_n\Pi^*_n\nonumber\\
&=&\Delta^2_n+\Omega^2_{n-1}+\Omega_{n-1}\Omega_{n}e^{i(\phi_{n-1}-\phi_{n}-\chi t)}\nonumber\\
&&+\Omega_{n-1}\Omega_{n+1}e^{i(\phi_{n-1}-\phi_{n+1}-2\chi t)}  \nonumber\\
&&+\Omega_{n}\Omega_{n-1}e^{i(\phi_{n}-\phi_{n-1}+\chi t)}+\Omega_{n}^2\nonumber\\
&&+\Omega_{n}\Omega_{n+1}e^{i(\phi_{n}-\phi_{n+1}-\chi t)}\nonumber\\
&&+\Omega_{n+1}\Omega_{n-1}e^{i(\phi_{n+1}-\phi_{n-1}+2\chi t)}\nonumber\\
&&+\Omega_{n+1}\Omega_{n}e^{i(\phi_{n+1}-\phi_{n}+\chi t)}+\Omega_{n+1}^2.
\end{eqnarray}
In the main text, we have chosen $\vec\theta=(0,-\pi/4,\pi/2)$ as an example. Assume that $\Omega_{n-1}=\Omega_{n}=\Omega_{n+1}=\Omega$ and $\Xi^2$ in different subspaces are as follows:
\begin{eqnarray}
&&\Xi_{n=0}^2(0,-\frac{\pi}{4},\frac{\pi}{2})=\Delta^2_0+\Omega^2[2+2\cos(\frac{\pi}{4}-\chi t)],\nonumber\\
&&\Xi_{n=1}^2(0,-\frac{\pi}{4},\frac{\pi}{2})=\Delta^2_1+\Omega^2[3+2\cos(\frac{\pi}{2}+2\chi t)],\nonumber\\
&&\Xi_{n=2}^2(0,-\frac{\pi}{4},\frac{\pi}{2})=\Delta^2_2+\Omega^2[2-2\cos(\frac{\pi}{4}-\chi t)].
\end{eqnarray}
Then, we use Pearson product-moment correlation coefficient to quantify the similarity of $\Xi^2_n$ for different subspaces. The correlation coefficient is
\begin{eqnarray}
    \rho_{X,Y}=\frac{E(XY)-E(X)E(Y)}{\sqrt{E(X^2)-(E(X))^2}\sqrt{E(Y^2)-(E(Y))^2}},
\end{eqnarray}
where $E(X)=\int^\tau_0Xdt/\tau$ is expressed as the average value of X. In the definition of $\rho_{i,j}$, $X=\Xi^2_{n=i}$, and $Y=\Xi^2_{n=j}$, a correlation-coefficient matrix can be caulclated as
\begin{eqnarray}
    \rho(0,-\frac{\pi}{4},\frac{\pi}{2})=\left(
    \begin{array}{ccc}
         1 & & \\
         & 1 & \\
         &  & 1
    \end{array}\right).
\end{eqnarray}
The above matrix only indicates that $\rho_{i,j}=1$, signifying a strong linear correlation  between $\Xi^2_{n=i}$ and $\Xi^2_{n=j}$. The remaining matrix elements, which represent weaker correlations, can be calculated as values less than 1. For instance,
\begin{eqnarray}
    \rho_{0,2}(0,-\frac{\pi}{4},\frac{\pi}{2})=\frac{2(\chi\tau)^2-\sin(2\chi\tau)\chi\tau+4\sin^2(\chi\tau)}{2(\chi\tau)^2+\sin(2\chi\tau)\chi\tau-4\sin^2(\chi\tau)};\nonumber\\
\end{eqnarray}
i.e., regardless of the value of $\chi\tau$, the correlation coefficient is always less than 1. The different subspaces are not entirely linearly related, which underscores the need to identify appropriate parameters that can satisfy the constraint relationships within each individual subspace.

From another perspective, the inclusion of counterrotating terms $(|m-n|\leq1)$ introduces varying degrees of perturbation to the constraint relationships, leading to a transition from regular similarity to increased disorder.

\section{Effect of higher-order terms}\label{highorder}
Our scheme can also be incorporated to compensate higher-order effects, ensuring that the performance of the system remains robust and the desired gate fidelity is preserved. For a qubit-oscillator dispersively coupled system, besides the linear terms in Eq. (\ref{H0}), there also exist higher-order terms, which include the self-Kerr term $\hat{H}_{Kerr}=-\hbar K\hat{a}^{\dagger2}\hat{a}^{2}/2$ and the second-order dispersive shift $\hat{H}_{\chi'}=\hbar\chi'\hat{a}^{\dagger2}\hat{a}^{2}|e\rangle\langle e|/2$. As an example, by choosing $K=2\pi\times3$ kHz and $\chi'=2\pi\times5$ kHz in the orange-slice-shaped scheme, we find that higher-order terms contribute to a reduction of only $0.21\%$ in the gate fidelity. This is considerably smaller than the $3.05\%$ decline caused by counterrotating terms and the $1.08\%$ reduction due to decoherence.

Nevertheless, the effect of the higher-order terms on the SNAP gate can be similarly analyzed by transforming them to the rotating frame with respect to the Hamiltonian $\hat{H}_0+\hat{H}_{Kerr}+\hat{H}_{\chi'}$. In this frame, resonant coupling is realized by selecting the drive frequency $\omega_d,m=\omega_q-\chi m+\chi'(m^2-m)/2$ and the transformed Hamiltonian reads
\begin{eqnarray}
\hat{H}'_{n,m}&=&\frac{\Delta_n }{2}\hat{\sigma}^z_n+\frac{\Omega_m(t)}{2}e^{i((n-m) \chi t+\phi_m(t))}\times\nonumber\\
    &&e^{-i\frac{1}{2}(n^2-n-(m^2-m)\chi't)}|g,n\rangle\langle e,n|+\mathrm{H.c.},
\end{eqnarray}
which is similar to the equation for the counterrotating terms. Thus, the impact of higher-order terms on the system evolution can also be eliminated, leading to more accurate gate operations.

The transmon qubit is also a suitable physical implementation candidate for our scheme. However, when independently controlling the subspace $\{|g,n\rangle,|e,n\rangle\}$, there may be simultaneous coupling to higher-lying states $|f,n\rangle$. Specifically, for a transmon qubit with anharmonicity $2\pi\times220$MHz, the contribution of such coupling to higher levels leads to only a 0.01$\%$ reduction in gate fidelity. Also, the derivative removal by adiabatic gate (DRAG) strategy would be a nice technique to prevent coupling to higher-lying states of the anharmonic qubit and our scheme is compatible with that technique.

Additionally, the contributions from higher-order counterrotating terms ($|m-n|>1$) can be mitigated by adjusting $\Pi_n$ to the form of
    $\Pi_n=\sum_{m,n}\Omega_{m}\exp[i((n-m)\chi t+\phi_m)]$.
Then, the relationship in Eq. (\ref{relationship}) will become more complex and finding an appropriate pulse shape to satisfy the constraint relationships will become more challenging. Nevertheless, it still provides valuable guidance on how to mitigate the effect of higher-order counterrotating terms.


\begin{thebibliography}{99}

\bibitem{Nielsen}  M. A. Nielsen and I. L. Chuang, Quantum Computation and Quantum Information, 10th ed. (Cambridge Univ. Press, Cambridge, 2010).


\bibitem{2012-surfacecode}
A. G. Fowler, M. Mariantoni, J. M. Martinis, and A. N. Cleland,
Surface codes: Towards practical large-scale quantum computation,
Phys. Rev. A \textbf{86}, 032324 (2012).

\bibitem{2023-surfacecode-StatePreparation}
Y. Ye, T. He, H.-L. Huang, Z. Wei, Y. Zhang, Y. Zhao, D. Wu, Q. Zhu, H. Guan, S. Cao,
{\it et al}.,
Logical Magic State Preparation with Fidelity beyond the Distillation Threshold on a Superconducting Quantum Processor,
Phys. Rev. Lett. \textbf{131}, 210603 (2023).

\bibitem{2023-surfacecode-25}
R. Acharya, I. Aleiner, R. Allen, T. I. Andersen, M. Ansmann, F. Arute, K. Arya, A. Asfaw, J. Atalaya, R. Babbush,
{\it et al}., Suppressing quantum errors by scaling a surface code logical qubit, Nature (London) \textbf{614}, 676 (2023).

\bibitem{GKP}
D. Gottesman, A. Kitaev, and J. Preskill,
Encoding a qubit in an oscillator,
Phys. Rev. A 64, 012310 (2001).

\bibitem{2021-bosoniccode-review}
W. Cai, Y. Ma, W. Wang, C.-L. Zou, and L. Sun,
Bosonic quantum error correction codes in superconducting quantum circuits,
Fundam. Res.  {\bf 1},  50 (2021). 


\bibitem{2021-bosoniccode-review2}
A. Joshi, K. Noh, and Y. Y Gao,
Quantum information processing with Bosonic qubits in circuit QED,
Quantum Sci. Technol \textbf{6}, 033001 (2021).

\bibitem{2024-GKP}
A. J. Brady, A. Eickbusch, S. Singh, J. Wu and Q. Zhuang,
Advances in bosonic quantum error correction with Gottesman¨CKitaev¨CPreskill Codes: Theory, engineering and applications,
Prog. Quant. Electron. \textbf{93}, 100496 (2024).


\bibitem{2016-catcode-lifetime}
N. Ofek, A. Petrenko, R. Heeres, P. Reinhold, Z. Leghtas, B. Vlastakis, Y. Liu, L. Frunzio, S. M. Girvin, L. Jiang,
{\it et al}.,
Extending the lifetime of a quantum bit with error correction in superconducting circuits,
Nature (London) \textbf{536}, 441 (2016).

\bibitem{2018-bosoniccode-review}
V. V. Albert, K. Noh, K. Duivenvoorden, D. J. Young, R. T. Brierley, P. Reinhold, C. Vuillot, L. Li, C. Shen, S. M. Girvin,
{\it et al}., Performance and structure of single-mode bosonic codes,
Phys. Rev. A \textbf{97}, 032346 (2018).


\bibitem{2019-Repetitioncatcode}
J. Guillaud and M. Mirrahimi,
Repetition Cat Qubits for Fault-Tolerant Quantum Computation,
Phys. Rev. X \textbf{9}, 041053 (2019).

\bibitem{2019-binomialcode}
L. Hu, Y. Ma, W. Cai, X. Mu, Y. Xu, W. Wang, Y. Wu, H. Wang, Y. P. Song, C.-L. Zou,
{\it et al}.,
Quantum error correction and universal gate set operation on a binomial bosonic logical qubit,
Nat. Phy. \textbf{15} 503 (2019).

\bibitem{2022-catcode}
C. Chamberland, K. Noh, P. Arrangoiz-Arriola, E. T. Campbell, C. T. Hann, J. Iverson, H. Putterman, T. C. Bohdanowicz, S. T. Flammia, A. Keller,
{\it et al}.,
Building a Fault-Tolerant Quantum Computer Using Concatenated Cat Codes,
PRX Quantum \textbf{3}, 010329 (2022).


\bibitem{2023-breakevenpoint}
Z. Ni, S. Li, X. Deng, Y. Cai, L. Zhang, W. Wang, Z.-B. Yang, H. Yu, F. Yan, S. Liu,
{\it et al}.,
Beating the break-even point with a discrete-variable-encoded logical qubit,
Nature (London) \textbf{616}, 56 (2023).

\bibitem{ion} D. Leibfried, R. Blatt, C. Monroe, and D. Wineland,
Quantum dynamics of single trapped ions,
Rev. Mod. Phys. \textbf{75}, 281 (2003).


\bibitem{cavity} J. Raimond, M. Brune, and S. Haroche,
Manipulating quantum entanglement with atoms and photons in a cavity,
Rev. Mod. Phys. \textbf{73},
565 (2001).

\bibitem{circuit} M. H. Devoret and R. J. Schoelkopf,
Superconducting Circuits for Quantum Information: An Outlook,
Science \textbf{339}, 1169 (2013).


\bibitem{2015-Uc}
S. Krastanov, V. V. Albert, C. Shen, C.-L. Zou, R. W. Heeres, B. Vlastakis, R. J. Schoelkopf, and L. Jiang,
Universal control of an oscillator with dispersive coupling to a qubit,
Phys. Rev. A \textbf{92}, 040303(R) (2015).

\bibitem{2015-SNAP}
R. W. Heeres, B. Vlastakis, E. Holland, S. Krastanov, V. V. Albert, L. Frunzio, L. Jiang, and R. J. Schoelkopf,
Cavity State Manipulation Using Photon-Number Selective Phase Gates,
Phys. Rev. Lett.  \textbf{115}, 137002 (2015).

\bibitem{2007-QOCFT}
J. Werschnik and E. K. U. Gross,
Quantum optimal control theory,
J. Phys. B 
\textbf{40} R175 (2007).

\bibitem{2006-QOC1}
D.M. Reich, M. Ndong, and C.P. Koch,
Monotonically convergent optimization in quantum control using Krotov's method,
J. Chem. Phys. \textbf{136}, 104103 (2012).

\bibitem{2013-QOC-QSL}
G. C. Hegerfeldt,
Driving at the Quantum Speed Limit: Optimal Control of a Two-Level System,
Phys. Rev. Lett. \textbf{111}, 260501 (2013).

\bibitem{2011-QOC-CRAB}
P. Doria, T. Calarco, and S. Montangero,
Optimal Control Technique for Many-Body Quantum Dynamics,
Phys. Rev. Lett. \textbf{106}, 190501 (2011).

\bibitem{2005-QOC-GRAPE}
N. Khaneja, T. Reiss, C. Kehlet, T. Schulte-Herbr\"{u}ggen,
and S. J. Glaser,
Optimal control of coupled spin dynamics: design of NMR pulse sequences by gradient ascent algorithms,
J. Magn. Res. \textbf{172}, 296 (2005).

\bibitem{2022-D-SNAP-o}
M. Kudra,  M. Kervinen, I. Strandberg, S. Ahmed, M. Scigliuzzo, A. Osman, D. P. Lozano, M. O. Thol\'{e}n, R. Borgani, D. B. Haviland,
{\it et al}.,
Robust Preparation of Wigner-Negative States with Optimized SNAP-Displacement Sequences,
PRX Quantum \textbf{3} 030301 (2022).

\bibitem{2017-Uc-o}
R. W. Heeres, P. Reinhold, N. Ofek, L. Frunzio, L. Jiang,
M. H. Devoret, and R. J. Schoelkopf,
Implementing a universal gate set on a logical qubit encoded in an oscillator,
Nat. Commun. \textbf{8}, 94 (2017).


\bibitem{AA} Y. Aharonov and J. Anandan,
Phase Change During a Cyclic Quantum Evolution,
Phys. Rev. Lett. {\bf 58}, 1593 (1987).


\bibitem{ngqc}    Y. Xu, Z. Hua, T. Chen, X. Pan, X. Li, J. Han, W. Cai, Y. Ma, H. Wang, Y. P. Song, Z.-Y. Xue, and L. Sun,
Experimental implementation of universal nonadiabatic geometric quantum gates in a superconducting circuit,
      Phys. Rev. Lett. {\bf 124}, 230503 (2020).

\bibitem{2023-Fast}
J. Landgraf, C. Fl\"{u}hmann, T. F\"{o}sel, F. Marquardt, and R. J. Schoelkopf,
Fast quantum control of cavities using an improved protocol without coherent errors,
Phys. Rev. Lett. {\b 133}, 260802 (2024).

\bibitem{2001-QOCFT-J1}
G. Turinici and H. Rabitz,
Quantum wavefunction controllability,
Chem. Phys. \textbf{267}, 1 (2001)

\bibitem{1988-QOCFT-J2}
A. P. Peirce, M. A. Dahleh, and H. Rabitz,
Optimal control of quantum-mechanical systems: Existence, numerical approximation, and applications,
Phys. Rev. A \textbf{37}, 4950 (1988).

\bibitem{1998-QOCFT-J2}
W. Zhu, J. Botina, and H. Rabitz,
Rapidly convergent iteration methods for quantum optimal control of population,
J. Chem. Phys. \textbf{108}, 1953 (1998).

\bibitem{2013-C-ka}
M. Reagor, H. Paik, G. Catelani, L. Sun, C. Axline, E. Holland, I. M. Pop, N. A. Masluk, T. Brecht, L. Frunzio {\it et al}.,
Reaching 10 ms single photon lifetimes for superconducting aluminum cavities,
Appl. Phys. Lett. \textbf{102}, 192604 (2013).

\bibitem{NGQC-SP2-O*}
C.-Y. Ding, L.-N. Ji, T. Chen, and Z.-Y. Xue,
Path-optimized nonadiabatic geometric quantum computation on superconducting qubits,
Quantum Sci. Technol. {\bf 7}, 015012 (2022).


\bibitem{2024-Circle}
Y. Liang and Z.-Y. Xue,
Nonadiabatic geometric quantum gates with on-demand trajectories,
Phys. Rev. Appl. \textbf{21}, 064048 (2024).

\bibitem{2018-PTC-1}
S. A. Caldwell, N. Didier, C. A. Ryan, E. A. Sete, A. Hudson, P. Karalekas, R. Manenti, M. P. da Silva, R. Sinclair, E. Acala {\it et al}.,
Parametrically Activated Entangling Gates Using Transmon Qubits,
Phys. Rev. Appl. \textbf{10}, 034050 (2018).

\bibitem{2018-PTC-2}
X. Li, Y. Ma, J. Han, T. Chen, Y. Xu, W. Cai, H. Wang, Y.-P. Song, Z.-Y. Xue, Z.-Q. Yin, and L. Sun,
Perfect Quantum State Transfer in a Superconducting Qubit Chain with Parametrically Tunable Couplings,
Phys. Rev. Appl. \textbf{10}, 054009 (2018).

\end{thebibliography}
\end{document}